\documentclass{emulateapj}

\input epsf.sty

\newcommand{\Msun}{M$_{\odot}$}

\def\gsim{\mathrel{\rlap{\lower 4pt \hbox{\hskip 1pt $\sim$}}\raise 1pt
\hbox {$>$}}}
\def\lsim{\mathrel{\rlap{\lower 4pt \hbox{\hskip 1pt $\sim$}}\raise 1pt
\hbox {$<$}}}

\begin{document}

\title{Subaru and Keck Observations of \\
the Peculiar Type Ia Supernova 2006gz at Late Phases\altaffilmark{1}}

\author{
K.~Maeda\altaffilmark{2,3}, 
K.~Kawabata\altaffilmark{4},
W.~Li\altaffilmark{5}, 
M.~Tanaka\altaffilmark{6},
P.~A.~Mazzali\altaffilmark{3,7,8}, \\
T.~Hattori\altaffilmark{9}, 
K.~Nomoto\altaffilmark{2,6}, and
A.~V.~Filippenko\altaffilmark{5}}

\altaffiltext{1}{
Based on data collected at the Subaru Telescope (operated by the 
National Astronomical Observatory of Japan) and the 
W. M. Keck Observatory (operated as a scientific partnership 
among the California Institute of Technology, the
University of California, and NASA; supported by the 
W. M. Keck Foundation).}
\altaffiltext{2}{Institute for the Physics and Mathematics of the 
Universe (IPMU), University of Tokyo, 
Kashiwano-ha 5-1-5, Kashiwa-shi, Chiba 277-8582, Japan; 
keiichi.maeda@ipmu.jp .}
\altaffiltext{3}{Max-Planck-Institut f\"ur Astrophysik, 
Karl-Schwarzschild-Stra{\ss}e 1, 85741 Garching, Germany.}
\altaffiltext{4}{Hiroshima Astrophysical Science Center, Hiroshima University, Hiroshima, Japan.}
\altaffiltext{5}{Department of Astronomy, University of California, 
Berkeley, CA 94720-3411.}
\altaffiltext{6}{Department of Astronomy, School of Science,
University of Tokyo, Bunkyo-ku, Tokyo 113-0033, Japan.}
\altaffiltext{7}{Instituto Nazionale di Astrofisica 
(INAF)-Osservatorio Astronomico di Padova, vicolo dell'Osservatorio, 5, 
I-35122 Padova, Italy.}
\altaffiltext{8}{Research Center for the Early Universe, School of
Science, University of Tokyo, Bunkyo-ku, Tokyo 113-0033, Japan.}
\altaffiltext{9}{Subaru Telescope, National Astronomical Observatory 
of Japan, Hilo, HI 96720.} 

\begin{abstract}
Recently, a few peculiar Type Ia supernovae (SNe) that show
exceptionally large peak luminosity have been discovered.  Their
luminosity requires more than 1~\Msun\ of $^{56}$Ni ejected during
the explosion, suggesting that they might have originated from
super-Chandrasekhar mass white dwarfs.  However, the nature of
these objects is not yet well understood.  In particular, no data have
been taken at late phases, about one year after the explosion. 
We report on Subaru and Keck optical spectroscopic
and photometric observations of the SN Ia 2006gz, which had been
classified as being one of these
``overluminous" SNe~Ia.  The late-time behavior is distinctly
different from that of normal SNe Ia, reinforcing the argument that SN
2006gz belongs to a different subclass than normal SNe~Ia.  However,
the peculiar features found at late times are not readily connected to
a large amount of $^{56}$Ni; the SN is faint, and it lacks [Fe~II] and
[Fe~III] emission. If the bulk of the radioactive energy escapes
the SN ejecta as visual light, as is the case in normal SNe~Ia, the
mass of $^{56}$Ni does not exceed $\sim 0.3$~\Msun. 
We discuss several possibilities to remedy the problem. With the
limited observations, however, we are unable to conclusively
identify which process is responsible.  An interesting possibility is
that the bulk of the emission might be shifted to longer wavelengths,
unlike the case in other SNe~Ia, which might be related to dense
C-rich regions as indicated by the early-phase data. Alternatively,
it might be the case that SN 2006gz, though peculiar, was actually 
not substantially overluminous at early times.
\end{abstract}


\section{INTRODUCTION}

There is general agreement that Type Ia supernovae (SNe Ia) are
thermonuclear explosions of white dwarfs (WDs; e.g., Nomoto,
Iwamoto, \& Kishimoto 1997; 
Hillebrandt \& Niemeyer 2000).  Thanks to the
uniformity of their light-curve shapes after applying a correction
factor (Phillips 1993; Phillips et al. 1999), they can be used as
``standard candles'' to measure cosmological parameters, which led to
the discovery of the accelerating expansion of the universe and dark
energy (Riess et al. 1998; Perlmutter et al. 1999; see Filippenko 2005
for a review). However, the nature of their progenitor systems has not
been resolved (e.g., Livio 2000; Hillebrandt \& Niemeyer 2000; Nomoto
et al. 2003), making it difficult to reliably predict potential
evolutionary effects that could add systematic errors to the
determination of cosmological parameters.

The progenitors of normal SNe Ia (Branch, Fisher, \& Nugent 1993; Li
et al. 2001) are believed to be WDs having nearly the Chandrasekhar
mass (hereafter Ch-SN Ia and Ch-WD).  The recent discovery of
extremely luminous SNe Ia raises the possibility that not all SNe Ia
originate from a single type of progenitor system.  Howell et
al. (2006) reported that SN Ia 2003fg (SNLS-03D3bb) reached an
absolute $V$-band magnitude of $M_V = -19.94$ ($H_0 = 70$ km s$^{-1}$
Mpc$^{-1}$, $\Omega_{\rm M} = 0.3$, flat universe).  Assuming that
this luminosity is powered by the decay chain $^{56}$Ni $\to$
$^{56}$Co $\to$ $^{56}$Fe as in other SNe Ia, they estimated $M_{\rm
56Ni} \approx 1.29$~\Msun\ (hereafter $M_{\rm 56Ni}$ is the mass of
$^{56}$Ni produced and ejected during the explosion).  Combining this
with other elements whose existence in the ejecta is evident from the
spectra, the ejecta and progenitor masses should exceed the
Chandrasekhar limit of a nonrotating WD. This was the first
observationally based suggestion of an SN Ia from a
super-Chandrasekhar WD (hereafter SupCh-SN Ia and SupCh-WD).  Since
then, two other possibly overluminous SNe~Ia have been reported: SN
2006gz (Hicken et al. 2007) and SN 2007if (Yuan et al. 2007).

\begin{deluxetable}{llll}
 \tabletypesize{\scriptsize}
 \tablecaption{Spectroscopy ($+ 341$ days)\tablenotemark{a}
 \label{tab:log_01}}
 \tablewidth{0pt}
 \tablehead{
 \colhead{Filter}
 & \colhead{Wavelengths}
 & \colhead{Airmass}
 & \colhead{Exposure (s)}
}
\startdata
 O58 + R300  & 5,800--10,200~\AA & 1.056 & 1200 \\
 None + B300 & 4,700--9,000~\AA  & 1.085 & 1200 \\
 Y47 + B300  & 3,800--7,200~\AA  & 1.165 & 1200 \\
\enddata
\tablenotetext{a}{The slit PA was 0 deg. The flux was calibrated with 
the standard star G191B2B (Massey \& Gronwall 1990). }
\end{deluxetable}

All of the available data for these SNe~Ia are only for the early
phases, $t \lsim 100$ d (hereafter $t$ is the time since the
explosion).  At later times, SNe Ia enter the nebular phase, when the
Fe-rich innermost ejecta, which are hidden at early phases, can be
directly observed (e.g., Axelrod 1988).  In this paper, we report
late-time spectroscopy and photometry of the 
potentially overluminous\footnote{In this paper, we define overluminous 
SNe Ia by the peak luminosity, corresponding to 
$M_{\rm 56Ni}$ (mass of $^{56}$Ni) 
$\gsim 1$ \Msun.
} 
SN 2006gz taken with the 8.2-m Subaru telescope and the 10-m Keck I 
telescope. In \S 2 we present the observations and data reduction.  Results 
are shown in \S 4, along with some comparisons to model calculations (\S
3).  SN 2006gz is intrinsically
different from normal SNe Ia.  However, our results are not readily
interpretable in the SupCh-SN Ia scenario, raising questions about the
nature of this new subclass of objects, as discussed in \S 5.

\section{OBSERVATIONS AND DATA REDUCTION}

Spectroscopy and imaging of SN 2006gz were performed on 2007 September
18 (UT dates are used throughout this paper) with the 8.2-m Subaru
telescope equipped with the Faint Object Camera and Spectrograph
(FOCAS; Kashikawa et al. 2002).  The epoch corresponds to $t = t_{\rm
peak} + 341$ d, where $t_{\rm peak}$ is the time at $B$-band maximum
(JD 2,454,020.2; Hicken et al. 2007) measured from the unknown
explosion date.  The field was also imaged on 2007 October 14 ($t =
t_{\rm peak} + 367$ d) with the $10$-m Keck I telescope equipped with
the Low Resolution Imaging Spectrometer (LRIS; Oke et al. 1995).  The
seeing conditions were excellent on both nights; star profiles had
full width at half-maximum intensity (FWHM) of $\sim 0\arcsec.8$ and
$0\arcsec.7$, respectively. The same field was imaged again on 2007
November 6 by the Subaru/FOCAS ($t = t_{\rm peak} + 390$ d), although
the seeing was not good (FWHM $\approx 1\arcsec.5$).

For the Subaru spectroscopy on September 18, we took three spectra
with exposure time 1200\,s each.  We used the $0\farcs 8$-wide slit
and the R300 grism with the O58 filter (wavelength coverage
5800--10200 \AA), the B300 grism with the Y47 filter (4700--9000 \AA),
and the B300 grism with no filter (3800--7200 \AA), in the three
separate exposures.  The standard star G191B2B (Massey \& Gronwall
1990) was also observed for flux calibration.  Although the
signal-to-noise ratio (S/N) is low, we succeeded in obtaining a
spectrum of the SN (\S 3). The spectroscopic observations are 
summarized in Table 1. 

\begin{deluxetable*}{llllll}
 \tabletypesize{\scriptsize}
 \tablecaption{Photometry
 \label{tab:log_02}}
 \tablewidth{0pt}
 \tablehead{
   \colhead{UT Date (2007)}
 & \colhead{Phase\tablenotemark{a}}
 & \colhead{Telescope}
 & \colhead{Exposure (s)}
 & \colhead{Band}
 & \colhead{Magnitude}
}
\startdata
September 18 & +341 & Subaru & 60   & $B$  & $ > 24.4 $ \\
            &      & Subaru & 60   & $V$  & $ > 24.2 $ \\
            &      & Subaru & 60   & $R$  & $ > 24.0 $ \\
October 14  & +367 & Keck   & 555  & $R$ & $ 25.5 \pm 0.3 $ \\ 
            &      & Keck   & 645  & $g$ & $ 25.5 \pm 0.2 $ \\
November 6  & +390 & Subaru & 1000 & $R$ & $ > 23.3 $ \\
\enddata
\tablenotetext{a}{Time since $B$-band maximum (days). }
\end{deluxetable*}

Subaru/FOCAS uses an atmospheric dispersion corrector (ADC). Its
performance is such that the chromatic elongation due to atmospheric
dispersion (Filippenko 1982) is less than $0.1''$ within the range
3500--11000 \AA\ at altitudes of 30--90$^\circ$, so the
atmospheric dispersion should be negligible regardless of the airmass
and the slit angle. We confirmed that the ADC worked correctly during
the observations; additionally, the airmass was low for both SN 2006gz 
(Table 1) and the standard star ($\sim 1.21$). Thus, though the slit
position angle of $0^\circ$ differed from the parallactic angle,
we believe that blue vignetting due to atmospheric dispersion
was negligible.

The total exposure time for imaging was 60\,s for each of $B$, $V$,
and $R$ on September 18, 555\,s for $R$ and 645\,s for $g$ on October
14, and 1000\,s for $R$ on November 6.  We obtained images of standard
stars (Landolt 1992) near SA98-634 on September 18 and around
PG0231+051 on November 6 for photometric calibration.  The 
imaging observations are summarized in Table 2.

In the September Subaru images the SN was not detected.  We obtained
an upper limit for the SN luminosity as follows.  First, the magnitude
$m_0$ which results in 1 photon count per second (ADU) was derived for
each band. Then we obtained the sky count $I_{\rm sky}$ (corresponding
to the sky magnitude $m_{\rm sky}$) and the dispersion ($\sigma_{\rm
sky}$) around the SN position.  
The $N\sigma$ magnitude at the detection limit is calculated as 
\begin{equation} m_{\rm lim} = -2.5 \log\left(N \frac{\sigma_{\rm sky}}{I_{\rm sky}}\right) + m_{\rm sky} \ .
\end{equation}
Adopting $3\sigma_{\rm sky}$ for $0\farcs 8\times 0\farcs 8$ binned images 
(i.e., the pixel size is nearly equal to the seeing), we estimated the limiting magnitude 
in each band as $m_B > 24.4$, $m_V > 24.2$, and $m_R > 24.0$ mag.

In October, the SN was marginally detected in both the $g$ and $R$
Keck images (Fig. 1).  
Since SN 2006gz was not readily identified in the late-time images, we
checked the position of the SN using the stacked $R$-band Keck image
(with a total exposure time of 555~s).  Astrometric
transformation between the discovery image of SN 2006gz taken with the
Katzman Automatic Imaging Telescope (KAIT; Filippenko et al. 2001) on
2006 September 26 and the combined LRIS image, using 8 stars near the
location of SN 2006gz, yields a precision of 0.2 pixel for the
position of the SN in the LRIS image.

\begin{figure}
\begin{center}
	\begin{minipage}[]{0.45\textwidth}
		\epsscale{1.0}
		\plotone{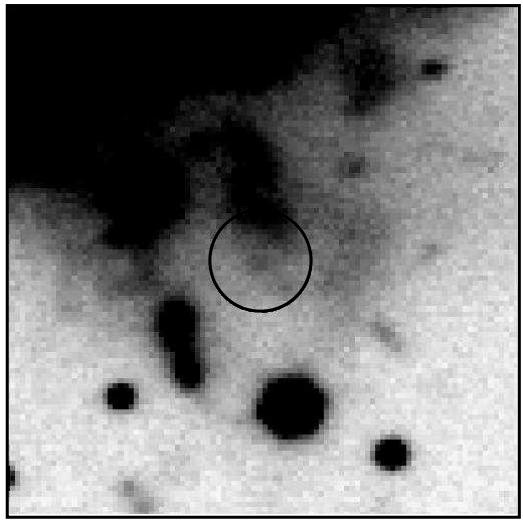}
	\end{minipage}
\end{center}
\caption
{A $21\arcsec.5 \times 21\arcsec.5$ section of the combined LRIS $R$-band image
centered on SN 2006gz taken with the 10-m Keck I telescope on 14 October 2007
(\S 2). North is up and east is to the left. The circle 
has a radius of 10 pixels, much larger than the uncertainty in the position
of SN 2006gz (0.2 pixel). 
\label{fig1}}
\end{figure}

Figure 1 shows a $21\arcsec.5 \times 21\arcsec.5$ section of the LRIS
image centered on SN 2006gz. The black circle has a radius of 10
pixels, which is much bigger than the astrometric transformation
uncertainty (0.2 pixel), and is used for clarity in the figure. At the
center of the circle, a faint stellar source is seen superimposed on
some extended background emission; we regard this as a detection of SN
2006gz.  Since the Subaru spectrum, although with poor S/N, shows
broad features especially at $\sim$7200--7300~\AA\ (\S 4.2), we conclude that
other possibilities, such as an unrelated background object or an
underlying H~II region, are unlikely.

The photometric reduction was performed using the IRAF package
DAOPHOT.  The SN was detected by the automatic star detection routine
DAOFIND, and then the photometry was performed via PSF fitting.  We
obtain the $R$ magnitude of the SN using the $R$ magnitude of field
stars derived from the September Subaru image.  For the $g$ magnitude,
the Subaru $B$ and $V$ images for the field stars are used, with
transformation equations from $B$ and $V$ to $g$ given by Fukugita et
al. (1996).  The photometry results in $m_g = 25.5 \pm 0.2$ and $m_R =
25.5 \pm 0.3$ mag.

In November the SN was not detected in the Subaru images.  Because of
the mediocre seeing, we did not obtain a meaningful $3\sigma$ limit:
$m_R > 23.3$ mag.

\begin{deluxetable}{lllllll}
 \tabletypesize{\scriptsize}
 \tablecaption{SN Ia Models\tablenotemark{a}
 \label{tab:model}}
 \tablewidth{0pt}
 \tablehead{
   \colhead{Name}
 & \colhead{$M_{\rm wd}$}
 & \colhead{$f_{ECE}$}
 & \colhead{$f_{\rm 56Ni}$}
 & \colhead{$f_{\rm IME}$}
 & \colhead{$M$($^{56}$Ni)}
 & \colhead{$E_{\rm k}$}
}
\startdata
SW7  & 1.39 & 0.18 & 0.43 & 0.32 & 0.6 & 1.40\\
LW7  & 1.39 & 0.00 & 0.72 & 0.21 & 1.0 & 1.41\\
SupCh2 & 2.00 & 0.21 & 0.50 & 0.22 & 1.0 & 1.60\\
SupCh3 & 3.00 & 0.14 & 0.33 & 0.46 & 1.0 & 1.50 
\enddata
\tablenotetext{a}{The units of mass and energy are 
\Msun and $10^{51}$ ergs, respectively. Also, 
$f_{\rm i}$ denotes the mass fraction of different 
burning products (\S 3). 
In all of the models, the mass fraction of unburned C+O 
is set to be 0.07.}
\end{deluxetable}

\section{SN Ia MODELS} 

To quantify the observational results, we have constructed four SN~Ia
models, based on the W7 model of Nomoto, Thielemann, \& Yokoi
(1984), which reproduces the basic observational features of normal
SNe~Ia (Branch et al. 1985).  Our model contains the following five
parameters (see also Howell et al. 2006; Jeffery et al. 
2006): $M_{\rm wd}$ (progenitor WD mass), $\rho_{\rm wd}$ (WD central
density, set to be $3 \times 10^9$ g cm$^{-3}$ throughout this paper),
$f_{\rm ECE}$ (mass fraction of electron capture Fe-peak elements,
e.g., $^{54}$Fe, $^{56}$Fe, $^{58}$Ni), $f_{\rm 56Ni}$ (mass fraction
of $^{56}$Ni), and $f_{\rm IME}$ (mass fraction of partially burned
intermediate-mass elements).  The kinetic energy of the ejecta
($E_{\rm K}$) is given as a function of these parameters, as this is
the nuclear energy generation reduced by the binding energy of the WD.
We use the binding energy formulae from Yoon \& Langer (2005), whose
models include rotating SupCh-WDs.  We adopt the density distribution
of the W7 model as a function of the velocity, and scale it in a
self-similar way: $\rho (v) \propto M_{\rm wd}^{5/2} E_{\rm
K}^{-3/2}$ and $v \propto M_{\rm wd}^{-1/2} E_{\rm K}^{1/2}$.  The
distribution of the elements is assumed to be concentric, and ordered
as follows: ECE, $^{56}$Ni, IME, then unburned C+O from the innermost
region.  Table 3 summarizes our models: SW7 (Simplified W7), LW7
(Luminous W7), SupCh2 and SupCh3 (Super-Chandrasekhar models), 
characterized by different
$M_{\rm wd}$ (progenitor WD mass), $M_{\rm 56Ni}$ (mass of $^{56}$Ni
synthesized at the explosion), and $E_{\rm K}$ (kinetic energy of the
expanding ejecta). 

\begin{figure}
\begin{center}
\hspace{-2cm}
	\begin{minipage}[]{0.55\textwidth}
		\epsscale{1.1}
		\plotone{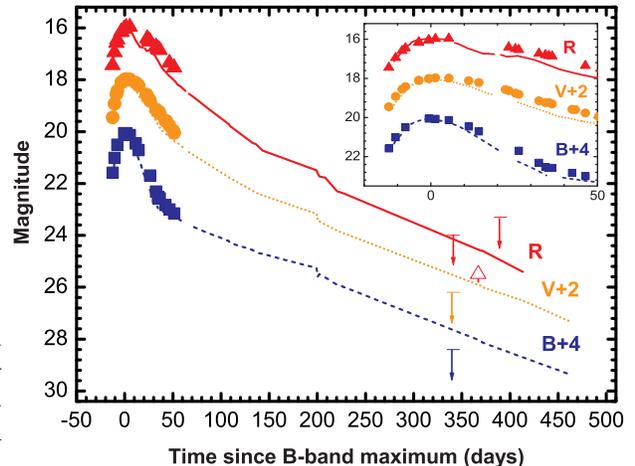}
	\end{minipage}
\end{center}
\caption
{$B$ (blue squares), $V$ (orange circles), and $R$ (red triangles)
light curves of SN 2006gz as compared with those of SN Ia 2003du ($B$
is the blue dashed line; $V$ is the orange dotted line; $R$ is the red
solid line). Each light curve of SN 2003du is arbitrarily shifted in
magnitude so that they are normalized at the peak and have roughly the
same peak magnitude as SN 2006gz ($\sim 16$ mag in all bands: Hicken
et al. 2007), to more easily compare the light-curve shapes.  SN
2003du represents a typical light-curve shape of SNe Ia (e.g., SN
2001el has an almost identical light-curve shape; Krisciunas, et
al. 2003; Stritzinger \& Sollerman 2007).  For SN 2006gz, the
early-phase data (filled symbols) are from Hicken et al. (2007), while
the late-phase date (open symbols) are from our Subaru (upper limit on
$+341$ d and $+390$ d) and Keck (open symbol, $+367$ d) observations.
SN 2003du data are from Stanishev et al. (2007).
\label{fig2}}
\end{figure}

For these models, we compute bolometric light curves using a one-dimensional
Monte-Carlo radiation transport code (Cappellaro et al. 1997; Maeda et
al. 2003) with the phenomenological opacity description for optical
photons ($\kappa_{\rm opt}$) given by Mazzali et al. (2001a), which
crudely takes into account the largest contribution from the Fe-peak
elements.  This description has been tested for normal SNe~Ia using
W7-like models (Mazzali et al. 2001a).  The transport of $\gamma$-rays
from the $^{56}$Ni/$^{56}$Co decays is treated in a gray approximation
with $\kappa_{\gamma} = 0.025$ cm$^{-2}$ g$^{-1}$, which is very
accurate for any input models (Sutherland \& Wheeler 1984; Maeda
2006).  Positron transport is also solved in a simplified way.  We
assume a phenomenological opacity $\kappa_{e^{+}} = 7$ cm$^2$
g$^{-1}$, a value that explains the slightly faster decline of
late-time light curves of normal SNe~Ia by increasing the fraction of
escaping positrons (Cappellaro et al. 1997).\footnote{Sollerman et
al. (2004) suggested that the light-curve behavior is explained by the
color evolution within the context of a fully trapped positron
scenario. The phenomenological opacity prescription here may therefore
overestimate the amount of positron escape.}

Synthetic nebular spectra are also computed at $t = t_{\rm peak} +
341$ d (i.e., Subaru/FOCAS observation in September), where $t_{\rm
peak}$ is obtained for each model by the light-curve calculations.
Given the deposited luminosity, which is obtained in the same way as
in the light-curve calculations, ionization-recombination equilibrium
and rate equations are solved iteratively (Mazzali et al. 2001b; see
also Maeda et al. 2006b). Since we have not
tried to obtain detailed fits to the observed data, the light curve and 
spectrum synthesis calculations
should be regarded only as indicative.

\begin{deluxetable*}{lllllll}
 \tabletypesize{\scriptsize}
 \tablecaption{Synthetic Magnitudes 
 \label{tab:phot}}
 \tablewidth{0pt}
 \tablehead{
   \colhead{Days since $B$ maximum}
 & \colhead{Band}
 & \colhead{SW7}
 & \colhead{LW7}
 & \colhead{SupCh2}
 & \colhead{SupCh3}
 & \colhead{Observed}
}
\startdata
+341 & $B$ & 24.5 & 23.9 & 23.7 & 23.2 & $>$ 24.4 ($\sim$ 24.8)\tablenotemark{a}\\
     & $V$ & 23.7 & 23.1 & 22.8 & 22.1 & $>$ 24.2 ($\sim$ 25.0)\tablenotemark{a}\\
     & $R$ & 24.8 & 24.3 & 23.9 & 23.0 & $>$ 24.0 ($\sim$ 25.0)\tablenotemark{a}\\
     & $m_{R} - m_{\rm Bol}$\tablenotemark{b}& 0.4  & 0.4  & 0.4  & 0.5 & \\
+367 & $R$  & 25.4 & 24.8 & 24.4 & 23.4 & 25.5 \\
     & $m_{R} - m_{\rm Bol}$\tablenotemark{b}& 0.5  & 0.4  & 0.5  & 0.5 & \\
\enddata
\tablenotetext{a}{Values in parentheses are the most probable
estimates, derived as follows.  For $+341$ d, $m_R$ is derived by
extrapolating the $R$-band magnitude at $+ 367$ d, assuming the
standard $^{56}$Ni/Co heating model.  Then the spectral flux was
calibrated with $m_R$, yielding an estimate of $m_B$ and $m_V$ at 
$+341$ d. Note that $m_B$ thus derived is highly uncertain, because 
of the very low S/N of the spectrum below 4,500~\AA. } 
\tablenotetext{b}{Derived for the synthetic model spectra.}
\end{deluxetable*}

\begin{figure}
\begin{center}
\hspace{-2cm}
	\begin{minipage}[]{0.55\textwidth}
		\epsscale{1.1}
		\plotone{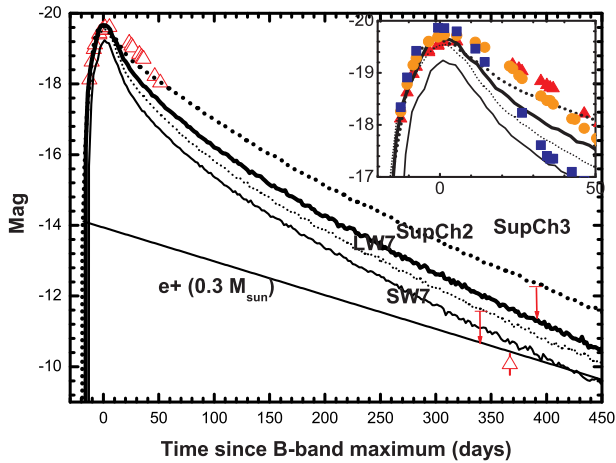}
	\end{minipage}
\end{center}
\caption
{Synthetic bolometric light curve for SN~Ia models, as compared with
the late-time $R$-band light curve of SN 2006gz (open triangles).
Note that $m_{R} - m_{\rm Bol} = 0.4$--0.5 mag for the models (\S 4.1
and Table 4).  Models shown here are SW7 (thin solid; $t_{\rm peak} =
17$~d), LW7 (thin dotted; $t_{\rm peak} = 20$~d), SupCh2 (thick solid;
$t_{\rm peak} = 20$~d), and SupCh3 (thick dotted; $t_{\rm peak} =
25$~d).  Inserted figure is for the early-phase monochromatic light
curves (Hicken et al. 2007), as compared with the same bolometric
model light curves.  We assume $\mu = 34.95$ mag, $E(B-V)_{\rm host} =
0.18$ mag, $E(B-V)_{\rm Gal} = 0.063$ mag, and $R_{V} = 3.1$ to 
convert the observed magnitudes to absolute magnitudes.
\label{fig3}}
\end{figure}

\section{RESULTS}

\subsection{Light Curve: Rapid Fading}

Figure 2 shows the results of our photometry as combined with the
early-phase light curve of Hicken et al. (2007). The early
post-maximum decline of SN 2006gz was slower in all bands than that of
normal SNe Ia (Hicken et al. 2007), represented here by SN 2003du.
However, this is not the case at the late phase: the detection of SN
2006gz in the $R$ band at $m_R = 25.5 \pm 0.3$ mag (Keck) and the
$3\sigma$ upper limit in $B$ and $V$ (Subaru) show that eventually the
visual luminosity of SN 2006gz declined more rapidly than that of other 
SNe~Ia. Note that SN 2003du has a typical late-time light-curve shape, with 
a decline rate of 1.5--1.6 mag (100~d)$^{-1}$. There 
have been several SNe~Ia whose decline is significantly slower (Lair 
et al. 2006), opposite to the behavior seen in SN 2006gz.

The September Subaru photometry ($+341$ d), $m_R > 24.0$ mag, is
consistent with the October Keck detection ($+367$ d) of SN 2006gz at
$m_R = 25.5 \pm 0.3$ mag.  The most likely magnitude at $t = t_{\rm
peak} + 341$ d is $m_R \approx 25.0$, $m_V \approx 25.0$, and $m_B
\approx 24.8$ mag\footnote{The $B$ magnitude here is highly uncertain,
because of the low S/N of the spectrum below $\sim$4500~\AA.},
assuming that the decline rate between these two epochs follows the
$^{56}$Ni heating model (see Table 4 for the model prediction).

Figure 3 shows the synthetic bolometric light curves of the four
models as compared with the observed $R$-band light curve.  Table 4
summarizes the synthetic multi-band magnitudes as derived from the
model spectra. The late-time bolometric correction is $m_{R} - m_{Bol}
= 0.4$--0.5 mag for all the models presented here.

\begin{figure}
\begin{center}
\hspace{-2cm}
	\begin{minipage}[]{0.55\textwidth}
		\epsscale{1.1}
		\plotone{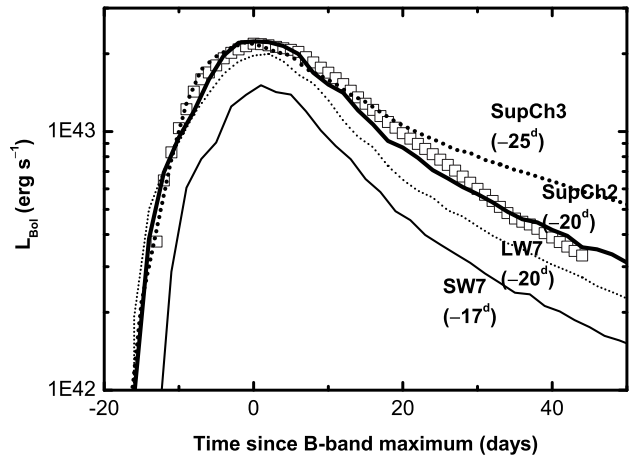}
	\end{minipage}
\end{center}
\caption
{Synthetic bolometric model light curves compared with the 
{\it UBVRI} bolometric light curve of SN 2006gz at early times 
(Hicken et al. 2007; see the footnote in the text). 
We assume $\mu = 34.95$ mag, $E(B-V)_{\rm host} =
0.18$ mag, $E(B-V)_{\rm Gal} = 0.063$ mag, and $R_{V} = 3.1$.
\label{fig4}}
\end{figure}

Figure 4 shows the synthetic bolometric light curves of the four
models as compared with the observationally derived bolometric light
curve of SN 2006gz [with $R_V = 3.1$, $E(B-V)_{\rm host} = 0.18$ mag, 
and $E(B-V)_{\rm Gal} = 0.063$ mag].\footnote{The
observationally derived bolometric light curve is presented at
http://www.cfa.harvard.edu/supernova/SNarchive.html (Hicken et
al. 2007).} The models with $M$($^{56}$Ni) 
= 1~\Msun\ reproduce the peak magnitude fairly well. The behavior
before the peak (except SW7) is very similar for different models
despite different $t_{\rm peak}$.  Thus, it seems difficult to
constrain $t_{\rm peak}$ as accurately as $t_{\rm peak} = 16.6 \pm
0.6$ d mentioned by Hicken et al. (2007), and our models are
consistent with the observed rising behavior.  The different
progenitor masses can be distinguished after the peak.  A more massive
progenitor results in a slower decline after the peak because of the
larger diffusion time scale. From this point of view, Model SupCh2
with $M_{\rm wd} = 2$~\Msun\ yields the light curve most consistent
with the observations around the peak.
In addition, the photospheric velocity and temperature at the peak luminosity,
derived in the light-curve calculations, are similar in the SupCh2
and SW7 models.  Within the uncertainties involved in the model
calculations, these are consistent with the observed characteristics.
Details of the early-phase characteristics will be presented elsewhere
(K. Maeda, in preparation).

At late epochs, however, the models are more than 1 mag brighter than
the observations, with the discrepancy reaching $\sim 2$\,mag for
SupCh2 and $\sim 3$\,mag for SupCh3 in the $V$ band, where the models
predict strong emission lines (see \S 4.2).  Indeed, only model SW7 is
marginally consistent with the late-time photometry, while the other
three models having $M_{\rm 56Ni} = 1$~\Msun\ are too bright compared
to the observations.

The failure of the SupCh models stems from the larger binding energy
of a WD and the higher density.  The peak date ($t_{\rm peak}$)
measured from the explosion date and the peak luminosity ($L_{\rm
peak}$) are roughly estimated by the following relations (e.g., Arnett
1982):
\begin{eqnarray}
t_{\rm peak} & \approx & 11 \ 
\left(\frac{\kappa_{\rm opt}}{0.1 \ {\rm cm}^2 \ g^{-1}}\right)^{1/2} 
M_{{\rm wd}, \odot}^{3/4} E_{{\rm K},51}^{-1/4} \ {\rm d}, \\
L_{\rm peak} & \approx & 7.8 \times 10^{43} \  
M_{{\rm 56Ni}, \odot} \exp\left(\frac{-t_{\rm peak}}{8.8~ {\rm d}}\right) 
\ {\rm ergs} \ {\rm s}^{-1}  \ .
\end{eqnarray}
Here the subscripts $\odot$ and $51$ mean that these values 
are expressed in units of \Msun\ and $10^{51}$ ergs, respectively. 
The optical depth to $\gamma$-rays from the $^{56}$Co decay 
at late epochs ($t = t_{\rm tail}$) is expressed 
as follows (e.g., Clocchiatti \& Wheeler 1997; Maeda et al. 2003): 
\begin{equation}
\tau_{\gamma} \approx 1000\, M_{{\rm wd}, \odot}^2 E_{{\rm K}, 51}^{-1} 
\left(\frac{t_{\rm tail}}{{\rm d}}\right)^{-2}\ .
\end{equation} 
The luminosity at $t_{\rm tail}$ ($L_{\rm tail}$) is then estimated as 
\begin{eqnarray}
L_{\rm tail} & \approx & 1.3 \times 10^{43}\, {\rm ergs} \ {\rm s}^{-1} 
M_{{\rm 56Ni}, \odot} \nonumber \\
&  & \times (\tau_{\gamma} + 0.035 f_{e+}) 
\exp\left(\frac{-t_{\rm tail}}{113.5~ {\rm d}}\right) \ , 
\end{eqnarray} 
where the factor 0.035 accounts for the positron contribution to the
heating, with $f_{e+}$ being the fraction of positrons trapped within
the ejecta ($f_{e+} \approx 1$ in usual situations).  Using these
expressions and values listed in Table 3, the expected bolometric
magnitude difference between the peak and the tail ($t_{\rm tail} =
t_{\rm peak} + 341$ d) can be derived: $\Delta m_{\rm Bol} \approx
7.1$ mag for SW7 and LW7, and $6.5$ for SupCh2.  This comes from the
larger $t_{\rm peak}$ and the larger $\tau_{\gamma}$ in the SupCh2
model, as a result of the larger mass and binding energy of a WD
(i.e., larger ratios of $M_{\rm wd}^{2}/E_{\rm K}$ and $M_{\rm
wd}^{3}/E_{\rm K}$, in which $E_{\rm K}$ is reduced by the binding
energy).  This behavior, smaller $\Delta m_{\rm Bol}$ in the SupCh
models, is independent of the uncertainty in distance and reddening.

Other model-independent constraints on $M_{\rm 56Ni}$ can be obtained 
considering the positron channel. Omitting the $\tau_{\gamma}$ term 
from eq. (5), we obtain an upper limit on $M_{\rm 56Ni}$ 
from the requirement that this positron luminosity should be below the 
observed luminosity (Fig. 3): 
\begin{equation}
M_{\rm 56Ni} \lsim 0.3 {\rm M}_\odot f_{e+}^{-1} \ .
\end{equation}
Here we adopted $m_{R} - m_{\rm Bol} = 0.5$ mag, a typical value in
the model spectrum synthesis calculations.

\begin{figure}
\begin{center}
\hspace{-2cm}
	\begin{minipage}[]{0.55\textwidth}
		\epsscale{1.1}
		\plotone{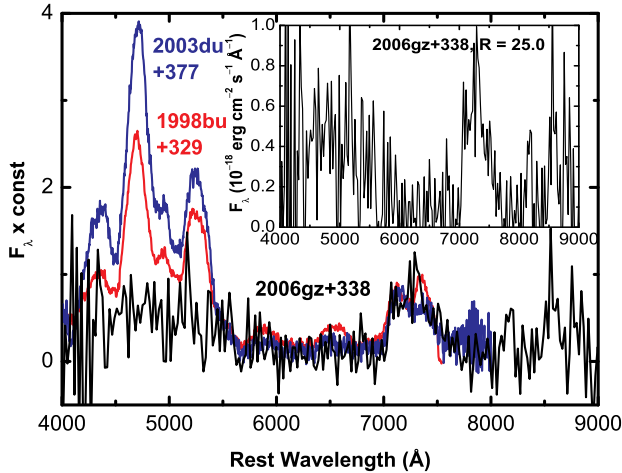}
	\end{minipage}
\end{center}
\caption
{Late-time Subaru spectrum of SN 2006gz at 
$t = t_{\rm p} +341$ d (black line) 
as compared with those of SNe Ia 2003du ($t_{\rm p} +377$ d, blue line; 
Stanishev et al. 2007) and 
1998bu ($t_{\rm p} +329$ d, green line; Cappellaro et al. 2001). 
In this comparison, the flux is normalized at $\sim$7000--7500~\AA. 
The insert is the Subaru spectrum, with the flux 
(in units of 10$^{-18}$ erg cm$^{-2}$ \AA$^{-1}$) 
renormalized to give $m_R = 25.0$ mag (see \S 4.1 and Table 4). 
Wavelengths are in the rest flame (using $z = 0.02363$ for SN 2006gz). 
The spectrum is smoothed with a binning width of 22.4~\AA. 
The data for SNe 2003du and 1998bu are from the SUSPECT database 
(see the footnote in the text).
\label{fig5}}
\end{figure}

\subsection{Spectrum: Missing [Fe~II] and [Fe~III] in the Blue}

Figure 5 shows the late-time spectrum of SN 2006gz and the comparison
with spectra of other SNe Ia.  In SN 2006gz, the emissions at
$\sim$4700~\AA\ ([Fe~II] $\lambda\lambda$4814, 4890, 4905; [Fe~III]
$\lambda\lambda$4858, 4701) and at $\sim$5200~\AA\ ([Fe~II]
$\lambda\lambda$5159, 5262) are extremely weak or undetected. The only
confirmed detection is at $\sim$7200--7300~\AA, which is interpreted
as the blend of [Fe~II] $\lambda\lambda$7151, 7171, 7388, 7452 in SNe
Ia. This feature may also be [Ca~II] $\lambda\lambda$7291, 7324 as
commonly seen in core-collapse SNe (see \S 5).

Indeed, in the SUSPECT database\footnote
{http://bruford.nhn.ou.edu/~suspect/index1.html .} we found only one
example of a SN Ia that probably shows a feature at
$\sim$7200--7300~\AA\ as strong as the [Fe~II] and [Fe~III] in the
blue. This is the underluminous SN 1991bg (Filippenko et al. 1992),
with $M_{\rm 56Ni} \approx 0.1$~\Msun (Mazzali et al. 1997).  Thus, SN
2006gz does not appear to follow the behavior of normal SNe Ia, both
in the light-curve shape (Hicken et al. 2007) and in the late-time
spectral features.

Figure 6 shows the synthetic spectra computed for the four models.
There is a tendency for more-massive models to show a weaker flux in
the [Fe~II]--[Fe~III] emission at $\lsim$5200~\AA\ relative to the
[Fe~II] line near 7200~\AA. An important quantity to characterize the
relative line strengths is the ratio of density to heating rate per
mass: generally, as this ratio increases, ionization (competing with
recombination) and temperature (competing with the line emission)
decrease, resulting in the stronger 7200~\AA\ feature (corresponding
to the lower ionization state and temperature). More-massive models
have the larger ratio (e.g., in SupCh2 the density is a factor of 2,
but the heating rate only a factor of 1.3, larger than in model SW7),
but it is not sufficiently large to reproduce the observed, extremely
large ratio of the red to the blue emission.

\begin{figure}
\begin{center}
\hspace{-2cm}
	\begin{minipage}[]{0.55\textwidth}
		\epsscale{1.1}
		\plotone{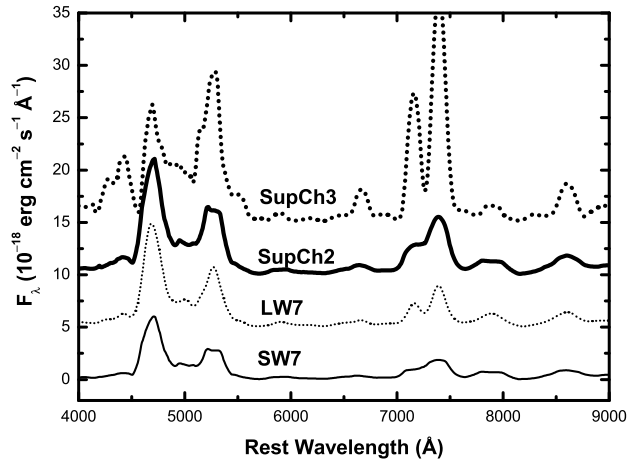}
	\end{minipage}
\end{center}
\caption
{Synthetic late-time spectra of the SN~Ia models. 
The model flux is converted to the observed flux using 
the distance modulus and reddening shown in the caption of Figure 3. 
For clarity, an arbitrary constant is added to the flux for each model 
($+5, +10$, and $+15 \times 10^{-18}$ erg cm$^{-2}$ s$^{-1}$ \AA$^{-1}$ 
for LW7, SupCh2, and SupCh3, respectively).
\label{fig6}}
\end{figure}

\section{DISCUSSION AND CONCLUSIONS}

We have presented late-time photometry and spectroscopy of SN 
2006gz obtained with the Subaru and Keck I telescopes.  These are the
first late-time observational data for an SN Ia that was claimed to be
overluminous at early phases.  Such SNe have been suggested to be the
explosions of SupCh-WD progenitors.

The spectrum is characterized by the weakness of the [Fe~II] and
[Fe~III] lines in the blue ($\lsim 5200$~\AA) relative to emission
lines in the red (either [Fe~II] or [Ca~II] at $\sim 7200$~\AA).  This
does not fit into the sequence of the late-time spectral behavior of
normal SNe Ia, confirming that SN 2006gz belongs to a different
subclass of objects than normal SNe Ia.  Coupled with this is the
problem of the observed faintness of SN 2006gz, especially in the $V$
band.

The SupCh2 model (with a 2~\Msun\ WD progenitor synthesizing 1~\Msun\
of $^{56}$Ni) is in good agreement with the early-phase bolometric
light curve (\S 4.1; but see \S 5.1), although it predicts a higher
late-time luminosity than the Ch-SN Ia models because of the larger
$M_{\rm 56Ni}$ and $\gamma$-ray deposition rate.  The luminosity of
Sup-Ch models exceeds that observed by more than 2 mag in the $V$
band. Furthermore, one derives $M_{\rm 56Ni} \lsim 0.3$~\Msun, under
the usual assumptions that the bulk of the deposited radioactive
luminosity emerges at visual wavelengths, and that positrons are
almost fully trapped within the ejecta.  In what follows, we list
possible solutions for this inconsistency.

\subsection{Reconsidering the Super Chandrasekhar Model?}

We should first mention that the peak luminosity of SN 2006gz derived
by Hicken et al. (2007) involves large uncertainty, as the host-galaxy
extinction is not well constrained.  If the host extinction is
negligible, the peak luminosity of SN 2006gz is close to that of
normal SNe~Ia ($M_{V} = -19.19$ mag).  In this case, SN 2006gz might
be a less energetic explosion of a Ch-WD, possibly because of
insufficient burning of the C-O layer to intermediate-mass elements,
to be consistent with the observed slow light-curve evolution at early
times.  It is not clear if the early-phase spectroscopic features can
be explained by this model: the early-phase spectra (Hicken et
al. 2007) indicate that the photospheric temperature is high (e.g.,
the weak Si II 5972) and the velocity is normal ($v_{\rm Si~II}
\approx 11,000$--12,000 km s$^{-1}$ at $B$-band maximum). Thus, the
most straightforward interpretation is that the peak luminosity is
also high.  Although the Ch-WD model is an interesting possibility and
deserves further investigation, it should not affect most of our
conclusions about the peculiar late-time behavior, as our arguments
are based mainly on the magnitude difference between the peak and the
tail. In what follows, we assume the host extinction adopted by Hicken
et al. (2007), but the above caveat should be kept in
mind.\footnote{For example, if the host extinction is smaller, then
the estimated values for $M_{\rm 56Ni}$ in the early and late-phases
both become smaller.}

From the late-time data, we find that $m_V$ is $\sim 1.3$ and $\sim
2.2$ mag larger (that is, fainter) than expected from the SW7 
($M_{\rm 56Ni} = 0.6$~\Msun) and SupCh2 ($M_{\rm 56Ni} = 1.0$~\Msun)
models, respectively.  This would imply that 0.1--0.2~\Msun\ of
$^{56}$Ni powers the late-time emission, smaller than the value
expected from the light-curve peak ($M_{\rm 56Ni} \approx 1$~\Msun)
by a factor of $\gsim 5$.  Such a large discrepancy cannot be
attributed solely to the effect of viewing angle, even if SN 2006gz
was a result of an extremely off-axis explosion (see, e.g., Sim et
al. 2007 for the effect of viewing angle; see also Maeda, Mazzali,
\& Nomoto 2006a).\footnote{This does not rule out the possibility that SN
2006gz is a Ch-SN Ia with a large offset and viewed from the side of
the $^{56}$Ni blob, as suggested by Hillebrandt et al. 
(2007) for another overluminous SN~Ia, SN 2003fg. 
Their model would have the same problem in reproducing the
late-phase data presented in this paper, and would require additional 
processes to explain it, as do the SupCh models.}

Alternatively, one may hypothesize that the energy source at early
times was not $^{56}$Ni/Co/Fe decay.  Strong circumstellar interaction
as seen in a peculiar class of SNe Ia (i.e., SNe Ia/IIa 2002ic-like
events: Hamuy et al. 2003; Deng et al. 2004; Wang et al. 2004) is not
favored, because of the lack of evidence for interaction in the
early-phase spectra.  Moreover, the SupCh models yield a reasonable
fit to the early-time {\it UBVRI} bolometric light curve without the
fine tuning of parameters.  The early-phase spectroscopic features
also seem to be consistent with SupCh models (\S 4.2; K. Maeda, in 
preparation). Thus, $^{56}$Ni/Co energy input is the most likely
mechanism to power the early-phase emission.

\subsection{Positron Escape, Infrared Catastrophe, or Dust Formation?}

If the early-phase emission is powered by $\sim 1$~\Msun\ of
$^{56}$Ni, then one is left with the possibility that some mechanism,
which is not at work in normal SNe~Ia, must be affecting the late-time
emission and the thermal conditions within the ejecta of SN 2006gz.

One possibility is the escape of positrons produced by the $^{56}$Co
decay out of the SN ejecta.  The issue has been comprehensively
explored by Milne, The, \& Leising (2001). They showed that a fraction
of the positrons can escape for a radially combed and/or weak magnetic
field, leading to the changing light curve at late times.  However, at
$t \approx 400$ d, this effect can account for at most $\sim 0.5$ mag,
which is not large enough to completely remedy the present problem.
 
Another possibility is the thermal catastrophe within the
$^{56}$Ni-rich region, which shifts the bulk of the emission from
optical to infrared wavelengths (Axelrod 1988). This so-called
``infrared catastrophe'' (IRC) is expected to take place after the
temperature drops below a few 1000~K, depending on the electron
density.  Observationally, the IRC has not been clearly detected in
any SNe Ia.  The IRC does not take place at $t \approx 1$ yr in any of
our models: the heating rate per unit volume in the Fe-rich emission
region is only a factor of $\sim 1.5$ smaller in SupCh2 than in SW7,
and thus the thermal conditions are similar in the two models.
 
Finally, it may be possible that the visual light from the
$^{56}$Ni-rich region is converted to near-IR/mid-IR wavelengths by
dust formed in the C+O-rich region. Before maximum light, SN 2006gz
showed the strongest C lines of any SN~Ia ever observed, with little
evolution in the absorption velocity. This indicates that a dense
C+O-rich shell or clumps are present in SN 2006gz, at least in the
outermost region.\footnote{Khokhlov, M\"{u}ller, \& H\"{o}flich (1993)
assumed that explosive carbon burning is ignited at the center of the
sub-Chandrasekhar mass ($\sim 1.2~ M_\odot$) degenerate C+O core (with
a central density as low as $\rho_c \approx 2 \times 10^8$ g
cm$^{-3}$) surrounded by a spherical extended envelope.  For such a
configuration, they demonstrated that interaction between the ejecta
and the hypothesized spherical envelope can give rise to a dense
shell.  However, the WD should evolve until the central density
becomes sufficiently high ($\rho_c \approx 2 \times 10^9$ g cm$^{-3}$)
for explosive carbon burning to occur.  At this point, the whole mass
would be in an almost hydrostatic rotating WD (Uenishi, Nomoto, \&
Hachisu 2003; Yoon \& Langer 2004) surrounded by a geometrically thin
accretion disk.}  The presence of carbon may be a key to the
understanding of the peculiar behavior at late phases, since carbon
has a relatively high condensation temperature.

The W7 model has $n \approx 10^7$ cm$^{-3}$ at the inner edge of the
C+O-rich region\footnote{ This relatively dense region is formed at
the interface between burning and non-burning layers. The density
structure is almost frozen because the time scale of the
Rayleigh-Taylor instability becomes longer than the time scale of the
bulk expansion of the WD. Such a density enhancement at the
(non-spherical) interface is thus also expected in multi-dimensional
explosions.} at 100~d after the explosion.  The corresponding
density in the SupCh models is a few times higher. Recently, Nozawa et
al. (2008) theoretically investigated dust formation in SNe~Ib.  They
found a large amount of carbon grain formation in the C-enhanced outer
He region, with a density comparable to that in the SN~Ia models, in their
model at $t \gsim 50$ d. Thus, carbon grain formation in SNe~Ia may
also be possible if carbon is abundant in the outermost region.  For
normal SNe~Ia, accelerated fading like that of SN 2006gz has never been
reported.  This may be consistent with the estimate that the upper
limit of the carbon mass fraction is as low as 0.01 (e.g., Tanaka et
al. 2008).  Details of the dust formation process, addressing the 
difference between SN Ib and SN Ia models (heating rate by
$^{56}$Ni, different oxygen mass fraction, etc.), should be investigated, 
as well as details of expected observational signatures during the 
dust formation. In this interpretation, a part of the outer Ca-rich 
region might be mixed with the C+O-rich region. It may partly explain 
the relative strength of the feature at $\sim 7300$~\AA\ relative to 
the [Fe~II] and [Fe~III] in the blue, since [Ca~II] arising from 
the Ca-rich region could be less diluted than the Fe emission, 
contributing to the 7300~\AA\ feature.

\subsection{Future Observations}

The possible interpretations we listed above are only speculative, but
they predict distinct observational signatures, to be tested in future
overluminous SNe~Ia.  First, late-time spectroscopy of other
potentially overluminous SNe~Ia would be extremely useful to see
whether SN 2006gz is special even among these peculiar SNe~Ia.  If
some of them show the SN 2006gz-like peculiarities at late phases, the
optical light curves spanning from early to late times should provide
a probe: (a) the IRC and dust scenarios predict a sudden decrease in
the visual luminosity, (b) the positron-escape scenario would manifest
itself with a relatively gentle decrease of the luminosity, and (c)
other heating scenarios would not necessarily follow the
quasi-exponential decline.

Near-IR light curves could directly show the effects of the IRC and
dust distinctly from other scenarios: we expect that a rapid increase
in the near-IR brightness should occur simultaneously with a rapid
decrease in the visual brightness, while other scenarios do not
necessarily predict this behavior.

A temporal sequence of optical spectra would also be highly useful.
In the dust formation scenario, we may see a transient red continuum
and emission-line shifts at intermediate phases, as was seen in the
peculiar SN Ib 2006jc (Smith, Foley, \& Filippenko 2008).

Finally, the IRC and dust scenarios could be unambiguously
distinguished with near-IR through mid-IR spectra. Blackbody radiation
from the dust particles should be seen in the dust scenario, while
line emission is the dominant cooling process in the IRC scenario.

\acknowledgements

We thank the director of Subaru, Masahiko Hayashi, as well as Hiroshi Terada,
for kindly allowing us to exchange observing time and targets in the
Subaru proposal S07B-057 (PI: K.M.), making the SN 2006gz observations
possible. The staffs at the Subaru and Keck Observatories are
acknowledged for their excellent assistance; moreover, Jeff Silverman,
Ryan Foley, and Maryam Modjaz helped with the Keck observations.  We
are also grateful to Stefan Taubenberger, Wolfgang Hillebrandt, Fritz
R\"opke, Daniel Sauer, and Stuart Sim for useful discussions.  This
research is supported by World Premier International Research Center
Initiative (WPI Initiative), MEXT, Japan, and by the Grant-in-Aid for
Scientific Research of the JSPS (18104003, 18540231, 20540226, 20840007) 
and MEXT (19047004, 20040004).  M.T. is supported through the JSPS (Japan
Society for the Promotion of Science) Research Fellowship for Young
Scientists. A.V.F.'s supernova group at the University of California,
Berkeley is grateful for the financial support of National Science
Foundation (NSF) grant AST--0607485 and the TABASGO Foundation. KAIT
and its ongoing research were made possible by donations from Sun
Microsystems, Inc., the Hewlett-Packard Company, AutoScope
Corporation, Lick Observatory, the NSF, the University of California,
the Sylvia \& Jim Katzman Foundation, and the TABASGO Foundation.
This research has made use of the CfA Supernova Archive, which is
funded in part by the NSF through grant AST--0606772.


\end{document}